\documentclass{article}
\newcommand{\bfr}{\begin{flushright}}
\newcommand{\efr}{\end{flushright}}
 
\begin{document}
\title{Finite Temperature Effect on Wilson Loop Mechanism
}
\author{Kiyoshi Shiraishi\\
Department of Physics, Tokyo Metropolitan University,
Tokyo 158
}
\date{Prog. Theor. Phys. Vol. {\bf 78} No. 3 (1987) pp. 535--539, 
(Progress Letters),  and Vol. {\bf 81} No. 1 (1989) pp. 248--248
(Erratum) }
\maketitle
\begin{abstract}
We evaluate the energy splitting of vacua appearing in the gauge
theory in the space $M_4\times S^N/Z_2$ ($N=2,3,4,5,6$ and $7$).
One-loop quantum effects which come from scalar and gauge fields are
considered. We calculate them at zero temperature as well as in
high-temperature limit. We find that in these situations there is no
breakdown of the gauge symmetry.
\end{abstract}

One of the interesting features of the heterotic string theory \cite{1}
has been based on the existence of gauge symmetries which can lead to
conceivable phenomenology at low energy.\cite{2,3} The mechanism
to break the symmetry is provided by the compactification,\cite{2,4}
which is another feature of the theory formulated in higher dimensions.
If the extra space is given by a {\it manifold} (not an orbifold
\cite{5}), the symmetry breaking is accountable in the region of field
theory, or low-energy effective theory of strings.
 
A few years ago Hosotani \cite{6} has shown the symmetry breaking
mechanism in the space-time $M_3\times S^1$, where $M_3$ denotes a
three-dimensional flat space-time and $S^1$ is a circle. In his model,
non-zero expectation value of the gauge field on $S^1$ affects the mass
spectra in $M_3$. At the same time those gauge fields leave the field
strengths to be zero. He considered one-loop quantum effects of gauge
and matter fields in order to find the correct vacuum because the energy
of each vacuum is degenerate at tree level.
 
Generally, similar symmetry breaking mechanism is to work when the
extra space is expressed by a multiply connected manifold, which is
often considered in the string theories.\cite{2,3} On such
manifolds, vacuum gauge fields $\langle A_m^a\rangle$, which satisfy
field strengths $\langle F^a_{mn}\rangle=0$, can be non-zero. To clarify
the issues, we may introduce Wilson loops associated with the gauge
fields:
\begin{equation}
U=P \exp\left(- i\int_\gamma \langle {\bf A}_m\rangle\,d{\bf x}^m\right)
\,,
\end{equation}
where $\gamma$ is the closed path on the non-simply connected space and
$P$ indicates that the products are taken as path-ordered. Non-trivial
Wilson line elements $U\ne 1$, which correspond to the non-contractible
loops on the manifold, may break the gauge symmetry. This is just the
way to break the large symmetry, which is expected to be crucial in
string theories.
 
 In general cases, the vacua associated with various Wilson elements
are degenerate at tree level. Evans and Ovrut \cite{7} took the one-loop
quantum effect into consideration to break the degeneracy of the
vacuum energy. They explained a way to calculate the vacuum energy in
Ref.~\cite{7} but did not show numerical results of the values.

On the other hand, we take an interest in the nature of this
symmetry-breaking mechanism in the cosmological context. It is widely
believed that the thermal effect gives rise to the symmetry restoration
or breaking in very early universe.\cite{8} For instance, usual Higgs
mechanisms are influenced by the finite-temperature effect and there
occur transitions between states which have different symmetries in
general.\cite{9} If the transition occurs also in the Wilson loop
mechanism, it may play an important role in the evolution of the
universe.
 
To investigate which symmetry realizes, we calculate the energy
splitting between the vacua. At the one-loop level finite temperature
effect is incorporated through the imaginary time formalism.\cite{9}
The high temperature and density effects on the Hosotani model which has
the compact space $S^1$ are investigated by the present
author.\cite{10} It has been concluded that the character of
symmetry is unchanged when temperature grows while it drastically
changes as the density of fermions grows.
 
In this paper we show numerical results of the energy gap between
vacua to the one-loop level on the compact manifold $S^N/Z_2$ at zero
temperature as well as in high temperature limit. Here $S^N/Z_2$ is a
quotient space constructed by identification of antipodal points on
$S^N$.\cite{7,11}
 
The harmonics on $S^N/Z_2$ are the same as those on $S^N$, but only
the modes which have ``reflection symmetry'' on $S^N$ are
permitted.\cite{7,11} For bosonic fields, the allowed harmonic
functions can be found from construction of harmonics on $S^N$ in
Ref.~\cite{12}.

Next we take the vacuum gauge field into account. If a gauge field (of a
certain symmetry) takes a nontrivial configuration, we can only prepare
the harmonics which change their sign under reflection defined on
$S^3$.\cite{7} For example, consider a scalar field on $S^N/Z_2$. On
$S^N$, the harmonics for a scalar function are labeled by integer $l$;
the eigenvalue of Laplacian is given by $-l(l+N-1)$. In the trivial
case,
${\bf A}_m=0$ on $S^N/Z_2$, the allowed modes are those defined on $S^N$
which is labeled by even $l$. When non-trivial configuration of gauge
fields exists on $S^N/Z_2$, odd-$l$ modes defined on $S^N$ appear in
some components of gauge multiplet.
 
Let us introduce the following quantity:
\[
\Delta F(\mbox{scalar})=F(\mbox{even})-F(\mbox{odd}) \,,
\]
where we consider the difference of energy (density) for a single
massless bosonic degree of freedom. $F(\mbox{even})$ and $F(\mbox{odd})$
are the vacuum free energy obtained from the one-loop calculation
adopting even-$l$ and odd-$l$ modes on $S^N$, respectively. The
difference of the energies of trivial and non-trivial gauge field vacua
is calculated to be $\Delta F$ multiplied by the number of components
which have the spectra of odd-$l$ modes. Thus it is necessary that
$\Delta F$ is positive for obtaining the non-trivial gauge vaccum with
lower free energy, although details about breaking patterns are
dependent on the gauge group.
 
The approach to evaluate $\Delta F$ at one-loop level is closely
related to the work of Candelas and Weinberg.\cite{11} For a scalar
field in the space-time $M_4\times S^N/Z_2$ with odd $N$, $\Delta F$
takes the following form at zero temperature:
\begin{equation}
\Delta F=-\frac{1}{2(4\pi)^{n/2}}\Gamma\left(-\frac{n}{2}\right)
\sum_{l=0}^\infty (-1)^{l}\frac{
\Gamma(l+N-1)}{l!\Gamma(N)}(2l+N-1)\cdot[l(l+N-1)]^{n/2}\,,
\label{eq2}
\end{equation}
where $n$ is to be set to four. Note that in this expression $(-1)^l$
indicates the difference between contributions from even and odd
modes. The calculation of $\Delta F$ is performed by a similar way to
Candelas and Weinberg's. Here we become aware that a straightforward
calculation provides a definite value of $\Delta F$ even for even $N$.

As is well known, for even dimensionality the expression of the vacuum
energy contains a logarithmic term with regularization mass scale at
the one-loop level.\cite{14} Especially for high temperature case, the
logarithmic term may cause some troubles. In this note, we believe that
$\Delta F$ is the expression (2) gives the general qualitative picture
even for even $N$.

At finite temperature $T=\beta^{-1}$, $\Delta F$ is written as follows:
\begin{equation}
\Delta F=-\frac{1}{2(4\pi)^{d/2}\beta}\Gamma\left(-\frac{d}{2}\right)
\sum_{n=-\infty}^\infty\sum_{l=0}^\infty
(-1)^{l}d_l(N)\left[l(l+N-1)+\left(\frac{2\pi}{\beta}n\right)^2\right]^{d/2}\,,
\label{eq3}
\end{equation}
where $d_{l}(N)=l(l+N-1)(2l+N-1)!/l!\Gamma(N)$ and $d$ is space
dimensionality to be set to three. As mentioned above, we may regard
(\ref{eq3}) as the definition of $\Delta F$ which stands even for even
$N$.
 
To illustrate the effect of temperature, we carry out the numerical
calculation in high temperature limit. In the limit of infinite
temperature, the expression of $\Delta F$ becomes simply as
\begin{equation}
\Delta F/T=-\frac{1}{2(4\pi)^{d/2}}\Gamma\left(-\frac{d}{2}\right)
\sum_{l=0}^\infty
(-1)^{l}d_l(N)\left[l(l+N-1)\right]^{d/2}\,,
\label{eq4}
\end{equation}
Note that at high temperature $\Delta F$ is proportional to $T$; this
dependence on temperature cannot be found by the usual method known as
high-temperature expansion,\cite{15} because the leading terms obtained
by the method are presented in terms of the total dimensionality and
curvatures.

\begin{table}[h]
\caption{The free energy splitting $\Delta F$ at zero
temperature obtained from the one-loop calculation of a scalar field
on $S^N/Z_2$, $N=2-7$.}
\label{t1}
\begin{center}
\begin{tabular}[t]{cc}
\hline
\hline
$N$ & $\Delta F$\\
\hline
$2$ & $-3.243\times 10^{-3}$\\
$3$ & $-5.116\times 10^{-3}$\\
$4$ & $-7.173\times 10^{-3}$\\
$5$ & $-9.404\times 10^{-3}$\\
$6$ & $-1.180\times 10^{-2}$\\
$7$ & $-1.435\times 10^{-2}$\\
\hline
\end{tabular}
\end{center}
\end{table}

\begin{table}[h]
\caption{$\Delta F/T$ in high temperature limit for a scalar field on
$S^N/Z_2$, $N=2-7$.}
\label{t2}
\begin{center}
\begin{tabular}[t]{cc}
\hline
\hline
$N$ & $\Delta F/T$\\
\hline
$2$ & $-1.18\times 10^{-2}$\\
$3$ & $-1.68\times 10^{-2}$\\
$4$ & $-2.19\times 10^{-2}$\\
$5$ & $-2.69\times 10^{-2}$\\
$6$ & $-3.21\times 10^{-2}$\\
$7$ & $-3.73\times 10^{-2}$\\
\hline
\end{tabular}
\end{center}
\end{table}

\begin{table}[h]
\caption{The free energy splitting $\Delta F$ at zero
temperature obtained from the one-loop calculation of a
transverse vector field on $S^N/Z_2$, $N=2-7$.}
\label{t3}
\begin{center}
\begin{tabular}[t]{cc}
\hline
\hline
$N$ & $\Delta F$\\
\hline
$2$ & $3.243\times 10^{-3}$\\
$3$ & $6.312\times 10^{-3}$\\
$4$ & $9.588\times 10^{-3}$\\
$5$ & $1.309\times 10^{-2}$\\
$6$ & $1.680\times 10^{-2}$\\
$7$ & $2.073\times 10^{-2}$\\
\hline
\end{tabular}
\end{center}
\end{table}

\begin{table}[h]
\caption{$\Delta F/T$ in high temperature limit for a
  transverse vector field on $S^N/Z_2$, $N=2-7$.}
\label{t4}
\begin{center}
\begin{tabular}[t]{cc}
\hline
\hline
$N$ & $\Delta F/T$\\
\hline
$2$ & $1.18\times 10^{-2}$\\
$3$ & $1.99\times 10^{-2}$\\
$4$ & $2.75\times 10^{-2}$\\
$5$ & $3.49\times 10^{-2}$\\
$6$ & $4.23\times 10^{-2}$\\
$7$ & $4.97\times 10^{-2}$\\
\hline
\end{tabular}
\end{center}
\end{table}

In Tables~\ref{t1} and \ref{t2} numerical results are given for
zero-temperature and high-temperature limit respectively. The
corresponding results for {\it transverse} vector fields are shown in
Tables
\ref{t3} and \ref{t4}. The full contribution of vector fields (which
contains the contribution of ghosts, scalars on $S^N/Z_2$) are obtained
by
$\Delta F(\mbox{transverse})+3\Delta F(\mbox{scalar})$ in the case with
$M_4
\times S^N/Z_2$.
 
It turns out that there is not case that non-zero gauge vacua have lower energy
than the ordinary vacuum for $N=2, 3, 4, 5, 6$ and $7$ at zero and
infinite temperature, as far as we consider scalar and vector fields
only. Judging from the tendency of the decrease of $\Delta F$ with $N$,
there is no gauge symmetry breakdown for arbitrary dimensionality $N$.
 
It seems that the temperature effect does not alter the determination
of the true gauge vaccum. We have checked the energy splitting at
arbitrary temperature for the space-time $M_4 \times S^3/Z_2$; in this
case, $\Delta F$ decreases smoothly with temperature and its sign is
unchanged.

In summary, for the space-time geometry $M_4\times S^N/Z_2$
($N=2,\dots,7$) we found no symmetry breaking caused by radiative
corrections which come from scalar and vector fields at zero temperature
and in high temperature limit. It Iooks plausible that phase
transitions do not take place by temperature effect in general Wilson
loop mechanism. This statement is supported by the fact that the
mechanism has no mass scale except for the scale of the compact
manifold.

If we desire the model with symmetry breaking, we must consider more
complicated manifold such as $S^3/Z_3$, and femionic fields.\cite{7}

In the above analysis we relied on the one-loop calculation of the
vaccum energy for a field theory. Recently compactified string theories
are studied by many authors.\cite{16,17,18} Narain \cite{16} showed that
the torus compactifications can bring about various gauge groups. This
mechanism is due to the non-zero gauge fields on tori.\cite{19} The
orbifold compactification \cite{5,17} which breaks the symmetry is
also well investigated because it is simple but reflects proper string
nature. We can ask whether the gauge vacuum with the lowest free energy
can change at finite temperature in these models. This subject may be
very important for constructing ``superstring cosmology''
scenario.\cite{20}

\bigskip

The author would like to thank T.~Hori for reading this manuscript. He
would also like to thank lwanami F\=ujukai for the financial support.

The author would like to thank J. S. Dowker and S. P. Jadhav for
pointing out the error in the first version of this paper.



\begin{thebibliography}{99}
\bibitem{1} D. Gross, J. Harvey, E. Martinec and R. Rohm, Phys. Rev.
Lett. {\bf 54} (1985) 502; Nucl. Phys. {\bf B256} (1985) 253; ibid.
{\bf B267} (1985) 75.
\bibitem{2} P. Candelas, G. Horowitz, A. Strominger and E. Witten,
Nucl. Phys. {\bf B256} (1985) 46.
\bibitem{3} J. Ellis, K. Enqvist, D. V. Nanopoulos and F. Zwirner,
Nucl. Phys. {\bf B276} (1986) 14, and references therein.
\bibitem{4} E. Witten, Nucl. Phys. {\bf B258} (1985) 75.

J. Breit, B. Ovrut and G. Segr\`e, Phys. Lett. {\bf B158} (1985) 33.

A. Sen, Phys. Rev. Lett. {\bf 55} (1985) 33.
\bibitem{5} L. Dixon, J. Harvey, C. Vafa and E. Witten, Nucl. Phys.
{\bf B264} (1985) 678; ibid. {\bf B274} (1986) 285.
\bibitem{6} Y. Hosotani, Phys. Lett. {\bf 126B} (1983) 309.
\bibitem{7} M. Evans and B. A. Ovrut, Phys. Lett. {\bf 174B} (1986) 63.

B. A. Ovrut, Prog. Theor. Phys. Suppl. {\bf 86} (1986) 185.
\bibitem{8} For example, see Nucl. Phys. {\bf B252} (1985) Nos. 1 \& 2.
\bibitem{9} D. A. Kirzhnits and A. D. Linde, Phys. Lett. {\bf B42}
(1972) 471.

L. Dolan and R. Jackiw, Phys. Rev. {\bf D9} (1974) 3320.

S. Weinberg, Phys. Rev. {\bf D9} (1974) 3357.

A. D. Linde, Rep. Prog. Phys. {\bf 42} (1979) 389.
\bibitem{10} K. Shiraishi, Z. Phys. {\bf C35} (1987) 37.
\bibitem{11} F. Ardalan and H. Arfaei, 
J. Math. Phys. {\bf 28} (1987) 685.
\bibitem{12} A. Chodos and E. Myers, Ann. of Phys. {\bf 156} (1984)
412.
\bibitem{13} P. Chandelas and S. Weinberg, Nucl. Phys. {\bf B237}
(1984) 412.
\bibitem{14} E. Myers, Phys. Rev. {\bf D33} (1986) 41.

R. Kantowski and K. A. Milton, Phys. Rev. {\bf D35} (1987) 397.
\bibitem{15} J. S. Dowker, Phys. Rev. {\bf D29} (1984), 2773; Class.
Quantum Grav. {\bf 1} (1984) 359.
\bibitem{16} K. S. Narain, Phys. Lett. {\bf 169B} (1986) 41.
\bibitem{17} L. E. Ibanez, H. P. Nilles and F. Quevedo, Phys. Lett.
{\bf B187} (1987) 25.

L. E. Ibanez, J. E. Kim, H. P. Nilles and F. Quevedo, Phys. Lett.
{\bf B191} (1987), 282.
\bibitem{18} H. Kawai, D. Lewellen and S.-H. Tye, Phys. Lett. {\bf
B191} (1987) 63.
\bibitem{19} P. Ginsparg, Phys. Rev. {\bf D35} (1987) 648.
\bibitem{20} M. Bowick and L. Wijewardhana, Phys. Rev. Lett. {\bf 54}
(1985) 2485.

M. Bowick, L. Smolin and L. Wijewardhana, Phys. Rev. Lett. {\bf 56}
(1986), 424.

N. Matsuo, Prog. Theor. Phys. {\bf 77} (1987) 223, and references cited
therein.
\end{thebibliography}
\end{document}